\documentclass[twocolumn,times,tighten]{aastex63}

\usepackage[utf8]{inputenc}
\usepackage{amsmath,amstext,amssymb}
\usepackage{rotating}
\usepackage{booktabs}
\usepackage{numprint}
\usepackage{graphics}
\usepackage{graphicx, float}
\usepackage{epsfig}
\usepackage{wrapfig}
\usepackage{xcolor}
\usepackage[figure,figure*]{hypcap}
\usepackage{natbib}

\definecolor{Green}{rgb}{0,0.5,0}
\definecolor{Blue}{rgb}{0,0,1}
\definecolor{Red}{rgb}{1,0,0}

\newcommand{\kms}{km s$^{-1}$}

\newcommand{\msun}{M$_{\odot}$}
\newcommand{\degrees}{\ensuremath{^\circ}}
\newcommand{\degree}{\degrees}

\shorttitle{Estimating the Masses of Supercluster-Scale Filaments from Redshift Dispersions} 
\shortauthors{Odekon et al.}

\begin{document}

\title{Estimating the Masses of Supercluster-Scale Filaments from Redshift Dispersions}

\author[0000-0003-0162-1012]{Mary Crone Odekon}
\affil{Department of Physics, Skidmore College, Saratoga Springs, NY 12866, USA; mcrone@skidmore.edu}

\author[0000-0001-9064-1132]{Trevor W. Viscardi}
\affil{Department of Physics, Skidmore College, Saratoga Springs, NY 12866, USA; twviscardi@gmail.com}

\author[0000-0003-0109-9308]{Jake Rabinowitz}
\affil{Department of Physics, Columbia University; jar2334@columbia.edu}

\author[0009-0001-9736-5596]{Brandon Young}
\affil{Department of Physics, Skidmore College, Saratoga Springs, NY 12866, USA; byoung@alumni.skidmore.edu}

\begin{abstract}
We present a strategy for estimating the mass per unit length along supercluster-scale filaments that are oriented across the sky, based on mock redshift surveys of 264 filaments from the Millennium simulation.  In our fiducial scenario, we place each simulated filament at a distance of 300 Mpc, perpendicular to the line of sight, and calculate the redshift dispersion using galaxies with magnitudes $r<19.5$. Some regions are dynamically complicated in ways that interfere with finding a simple relationship between dispersion $\sigma$ and linear mass density $\mu$.  
However, by examining individual overlapping segments along the filaments, we find a relationship that allows us to successfully predict $\log\mu$  from $\log\sigma$ with a scatter of about ± 0.20 dex, for $\sim$ 70\% of the regions along filaments. This relationship is robust to changes in the distance to the filament if the physical segment length and the absolute magnitude for galaxy selection are held constant. 
The relationship between redshift dispersion and mass is similar to that obtained for a simple analytical model where filaments are dynamically relaxed, and we examine the possibility that the galaxies are indeed relaxed within the gravitational potential of the filament. We find that this is not the case; galaxy dynamics are strongly effected by infall to the filament and by orbits within groups and clusters. 

\end{abstract}


\section{Introduction}
\label{sec:intro}
Over the past century, galaxy redshift surveys have provided an increasingly detailed and expansive view of the three-dimensional structure of the universe. By the 1990s, with the advent of multi-object spectrographs, tens of thousands of galaxy redshifts had been measured, confirming the existence of superclusters, voids, walls, and filaments (e.g. \citealt{delapparent1986a,haynes1986a}; see also the reviews by \citealt{geller1988a}, \citealt{giovanelli1991a}, and \citealt{strauss1995a}).
The properties of these structures present tests of cosmological models over a wide range of scales and dynamical environments.  

In addition to using redshift as a proxy for distance to map the positions of galaxies, it was recognized very early  \citep{zwicky1933a} that redshifts caused by peculiar motions could provide dynamical estimates of the underlying matter distribution itself. With the availability of large redshift surveys, \cite{eisenstein1997a} considered the possibility of extending this technique from individual clusters to large, supercluster-scale filaments. 
In particular, observations of filaments that are positioned across the sky, approximately perpendicular to the line of sight, offer the intriguing possibility of providing a dynamical estimate of linear mass density based simply on their width in redshift space. This method is even simpler than that for clusters of galaxies, in the sense that the mass may be found from the dispersion alone, without the need to include an estimated size like the radius that is needed for dynamical mass estimates of clusters. 

Using an approximation of thermal equilibrium within the gravitational potential of a long, uniform cylinder, \cite{eisenstein1997a} derived a relationship between the one-dimensional dispersion $\sigma$ perpendicular to the filament axis, and the mass per unit length $\mu$:  

\begin{equation}
    \mu = 7.4\times10^{13}M\textsubscript{\(\odot\)} \  \textrm{Mpc}^{-1} (\sigma/400 \ \textrm{\kms})^2
\end{equation} 

Recognizing that this simple analytical model does not take into account the more complicated structure and dynamics of actual filaments, \citet{eisenstein1997a} performed preliminary tests using N-body simulations. These tests, while limited by a mass resolution of $1.5\times10^{12}\ h^{-1}$ \msun \  and by including mock observations for only two of the simulated filaments, supported the possibility that velocity dispersion might be used to estimate mass in filaments oriented across the sky. 

Since the 1990s, new generations of cosmological simulations have vastly improved the prospects for understanding the expected relationship between filament redshift dispersion and the underlying mass. For example, \citet{pereyra2020a} compared perpendicular velocity dispersion in simulated filaments to their linear mass density using simulations with a mass resolution of $1.18\times10^{9}\ h^{-1}$ \msun. Using the full three-dimensional information for positions and velocities (not mock observations), they found results consistent with the relationship derived by \citet{eisenstein1997a}. They point out that this ``would suggest that the filaments are partially virialized in the direction perpendicular to the axis," while also noting that filaments are disturbed by infall and are therefore not fully relaxed.

Along with improvements in cosmological simulations, new generations of redshift surveys have vastly improved our ability to identify and study large-scale filaments in the actual universe. Indeed, large-scale filaments have attracted significant interest in recent years, primarily in the context of understanding the effect of large-scale structure on galaxy evolution. For example, galaxies in or near filaments tend to be more passive (e.g. have a lower specific star formation rate) and to have have less loose hydrogen gas \citep{alpaslan2014a, martinez2016a, malavasi2017a, chen2017a, odekon2018a, kraljic2018a, laigle2018a, salerno2019a, bonjean2020a, hoosain2024a}. In addition, the orientation of galaxy disks and angular momentum appears to be related to filament orientation \citep{jones2010a, tempel2013a,zhang2013a,chen2019a,kraljic2019a,sachin2022a}.  Studies like these show that galaxy properties depend not only on local density, but on their location within larger structures.

A different approach in the study of filaments, and one that is more similar to that used in this paper, is to focus on the overall structure of filaments themselves --- e.g. size and shape --- which can then be compared with theoretical expectations from cosmological simulations. Observational measures of filaments include length and galaxy overdensity \citep{rost2020a}, as well as large-scale galaxy infall patterns that reveal information about the filament mass distribution \citep{odekon2022a}. The latter paper is especially similar to our approach in that it relates galaxy motions to filament mass.  It differs, however, in that it requires redshift-independent distances to map out the peculiar velocity field surrounding each filament. Because redshift-independent distance are likely to have large uncertainties (uncertainties of $\sim30\%$ in the case of the baryonic Tully-Fisher distances modeled in \citealp{odekon2022a}), their strategy involves averaging over many galaxies to find an average infall profile.  This provides a mass estimate on a very large scale for the filament as a whole. The approach we present here is complementary in that it provides a way to find filament masses in smaller regions along the filament but, as we shall see, not for \textit{every} region of the filament.

As we completed this paper, \citet{zhu2024a} published a study finding a relationship between spatial filament width and linear halo mass density within cosmic filaments, based on hydrodynamic simulations. Their approach focuses on a detailed analysis of the  three-dimensional structure of filaments (as opposed to mock observations), and the way that embedded dark matter halos are related to the mass of filaments as a whole. Our paper provides a complementary analysis that focuses on finding a realistic observational strategy for determining filament masses from redshift surveys.

This paper is organized as follows. In Section 2 we describe our method for creating mock observations of filaments. In Section 3 we present a strategy for relating the observable redshift dispersions to the underlying dark matter distribution. In Section 4 we show that the relationship between dispersion and the matter is distribution does not imply dynamical equilibrium, but arises from a combination of infall and orbits within substructures, and in Section 5 we provide a summary and discussion.

In this paper, the dimensionless Hubble constant $h\equiv H_o/100$ km s$^{-1}$ Mpc$^{-1}$ is generally set to $h=0.7$. An exception is that masses and distances from the Millennium simulation are presented in terms of $h$ explicitly, the same way they are presented in the Millennium simulation database.
 
\section{Mock Observations}

We use the Millennium simulation \citep{springel2005a} to 
create mock observations of a large number of simulated filaments using realistic galaxy selection criteria. The Millennium simulation volume is a periodic cube of length 500 $h^{-1}$ Mpc, or about 714 Mpc for h=0.7. 
As described below, this volume is large enough that we can obtain a sample of 264 supercluster-scale filaments. For comparison, the Illustris TNG300 \citep{pillepich2018a} simulation volume is a cube of length 300 Mpc, nearly fifteen times smaller. The Millennium database also includes galaxy catalogs based on semi-analytic models. In this paper, we focus on the galaxy catalog of \citet{lagos2012a}, which applies the GALFORM model to the ``MR" run. This choice allows us to compare the galaxy velocity field to the large-scale matter distribution, which is provided in the Millennium database for the MR run (but not the later MR7 run) in the form of a $256^3$ density grid. At the time of this writing, we expect the MillenniumTNG \citep{hernandez-aguayo2023a} public database to be available in the near future. This will likely provide an excellent resource for follow-up work on supercluster-scale filaments. 

We define filaments with the same graph-based method used by \citet{odekon2022a}.  
The filaments are defined using groups of galaxies with virial masses above $10^{14} \ h^{-1}$ \msun \ (corresponding to $M_{200} > 0.72 \times 10^{14} \ h^{-1}$ \msun \ in the context of the simulation; see \citealp{saez2016a}). These are connected with a minimum spanning tree, which is then broken where connecting edges are greater than $17\ h^{-1}$ Mpc. The remaining structures that have at least six large groups are considered filaments. Breaking the tree at this particular edge length maximizes the number of filaments with at least six connected groups in the simulation. As shown in \citet{odekon2022a}, this method can be used to find large, supercluster-scale filaments in either observations or simulations. Applying this method to the Millennium MR run, we find 264 supercluster-scale filaments. 

For context, note that cosmological filaments are defined in a variety of ways, depending on the information available and the purpose of a particular study. Different methods for finding filaments are reviewed in \citet{libeskind2018a}, \citet{rost2020a}, and \citet{carronduque2021a}. The method we use here is similar to the ones used by \citet{alpaslan2014a}, \citet{odekon2018a}, \citet{pereyra2020a}, \citet{bonnaire2020a}, and other graph-based methods \citep[see section 2 of][]{libeskind2018a}.  

To create mock observations, we place each simulated filament across the sky at a distance of 300 Mpc. More specifically, we use principal component analysis applied to the large groups that were used to find the filament, with groups weighted in proportion to their mass. We then place the center of the principle axis of the filament 300 Mpc away, with the axis oriented perpendicular to the line of sight. 

Galaxies from the \citet{lagos2012a} semi-analytic catalog are chosen using a magnitude limit of $r<19.5$.  This magnitude limit is similar to that for several recent and upcoming wide-angle surveys such as the Galaxy and Mass Assembly survey (GAMA), which covers over 285 square degrees \citep{driver2022a}, and the Dark Energy Spectroscopic Instrument Bright Galaxy Survey (DESI BGS), which covers over 10000 square degrees \citep{hahn2023a}.  

Figure 1 shows the resulting mock observations and mass density distribution for one sample filament. On the left, grey dots indicate galaxies within a wedge of thickness $2\degrees$. 
On the right, color indicates density (from the $256^3$ counts-in-cell density grid in the Millennium Run MField database) within a thicker $4\degrees$ wedge to show the context for these galaxy motions in terms of the underlying mass distribution. In both the mock observation and the mass distribution, the location of the filament is clear, but its structure is much more complicated than a uniform cylinder. As we discuss below, this complicated structure presents a  challenge to finding a simple relationship between redshift dispersion and mass.  

\begin{figure*}
\includegraphics*[width=0.75\textwidth]{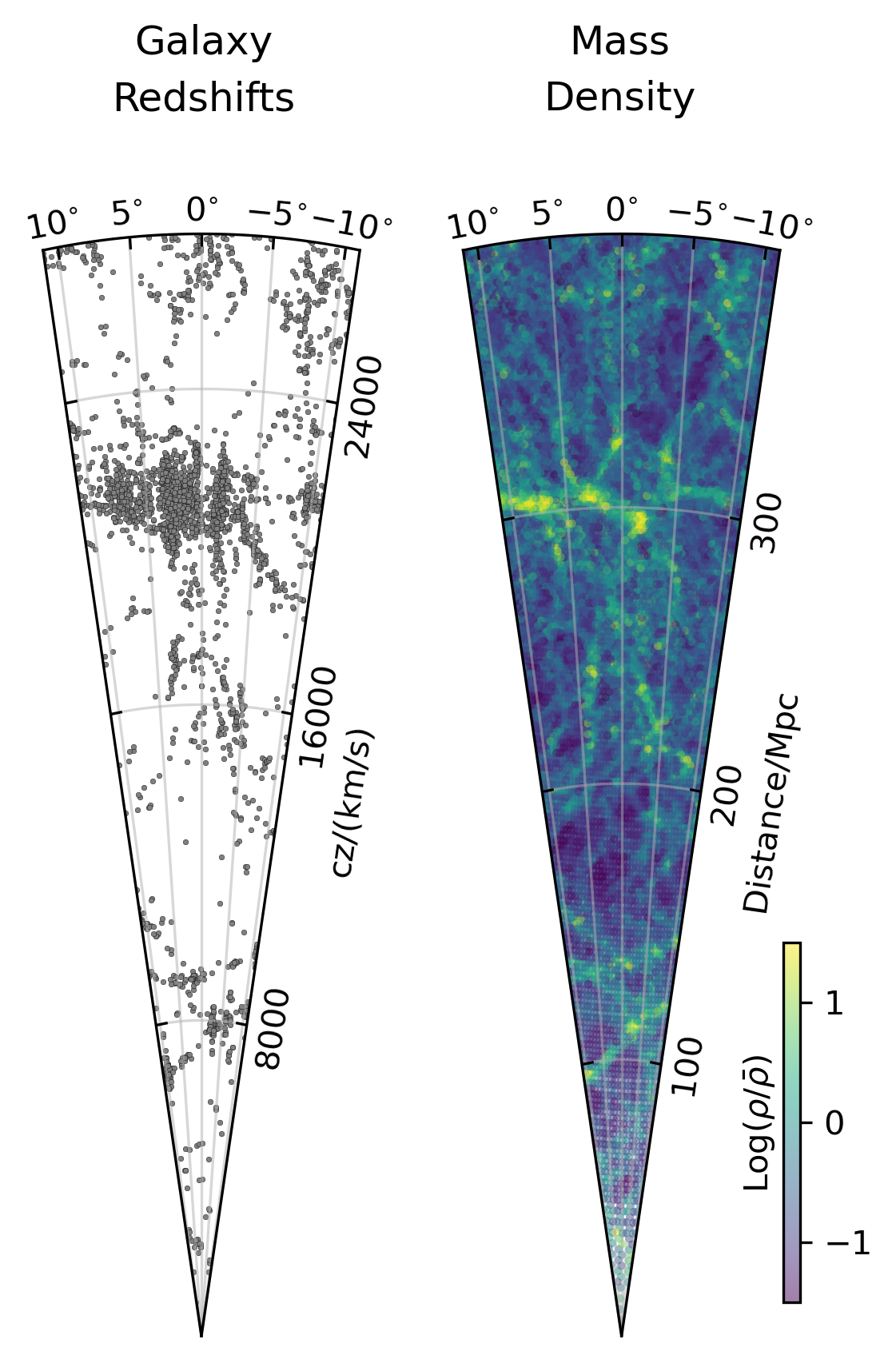}
    \caption{Mock observation (left) 
    and matter distribution (right) for one sample filament from the Millennium simulation. The filament is placed at a distance of 300 Mpc, corresponding to cz $\sim$ 21,000 km s$^{-1}$. 
    On the left, grey dots indicate the redshifts of galaxies with magnitude $r<19.5$ within a wedge of thickness 2\degrees. On the right, color indicates density, normalized to the average density of the universe, within a thicker 4\degree wedge to show the context for these motions in terms of the underlying mass distribution. In both diagrams, the location of the large filament is clear, but it has a complex structure and is embedded in a network of smaller filaments.}
\end{figure*}

\section{Relating Redshift Dispersion and Linear Mass Density }

\subsection{Determining Dispersion and Mass}

In this section, we describe a strategy for relating the observable redshift dispersion $\sigma$ of filaments, described in terms of the redshift multiplied by the speed of light $cz$, to their linear mass density $\mu$. Because supercluster-scale filaments are not uniform, our approach is to analyze each filament in individual segments. This allows us to find a more detailed picture of how the matter is distributed, and it allows us to discard regions of the filament that are too poorly-defined or dynamically complicated to provide a good mass estimate. 

For each filament, we examine redshifts in a region extending for $10\degrees$ along its principle axis ``longitude" and $2\degrees$ perpendicular to its principal axis (``latitude"). 
Individual segments have a length of $1\degrees$ along the filament, corresponding to about 5 Mpc for filaments 300 Mpc away. This segment length provides a balance between being large enough to yield an acceptable galaxy sample size, and small enough to avoid regions that are too poorly-defined or dynamically complicated. To maximize coverage, we use overlapping segments, where segment centers are shifted by 0.2\degree. The total number of segments for each filament is therefore 10\degrees/0.2\degrees = 50, and the total number segments for all 264 filaments is 13200. 

For each segment, we start by selecting values for $cz$ within the range 18000-24000 km s$^{-1}$, a window of width 6000 km s$^{-1}$ centered on the approximate location of the filament.  We then estimate the location and scale -- that is, the position and dispersion of the filament in redshift space. Several measures of location and scale have been studied in the context of velocity dispersions for clusters of galaxies \citep[e.g.][]{beers1990a}. While we expect the redshift distribution for filaments to be even more complicated than that for clusters in terms of substructure and infall, we face a similar motivation for estimators that are resistant to outliers, robust to assumptions about the underlying distribution, and efficient in the case of small sample size. With this context in mind, we consider three measures for the dispersion.  

\begin{itemize} 

\item{The \textbf{biweight} location and scale \citep{tukey1958a} have become standard measures in the context of clusters of galaxies, after \cite{beers1990a} found it to be ``consistently superior for most applications." While the biweight is already designed to be resistant to outliers, several authors have found it useful to additionally exclude outliers using sigma clipping \citep{yahil1977a} where galaxies are excluded if their redshifts are too far from the central location \citep[e.g.][]{wojtak2018a, ferragamo2020a, wetzell2022a}. In the context of filaments, we find outliers to be a significant challenge, especially for segments with few galaxies, where a large fraction of galaxies in the 6000 km s$^{-1}$ window around the filament might be considered outliers. We use fairly aggressive 2.5 sigma clipping, and even 2.0 sigma clipping to help with cases where the mass is low, as described in more detail below. We also use a smaller 2000 km s$^{-1}$ window for the first iteration of the calculation, before expanding to including the full 6000 km s$^{-1}$ to help select the region of the filament itself rather than an outlying aggregation of galaxies. Finally, we quantify the uncertainty as the standard deviation in values obtained through bootstrap resampling. 
}
\item{The \textbf{gapper} \citep{wainer1976a} is a measure of scale that is especially effective for small numbers of galaxies \citep{beers1990a, ferragamo2020a}. As with the biweight scale, we use the gapper iteratively with 2.5 sigma or 2.0 sigma clipping, begin the first iteration with a limited range of 2000 km s$^{-1}$ before expanding to include 6000 km s$^{-1}$, and quantify the uncertainty using bootstrap resampling. Because the gapper is a measure of scale only (not location), we use the biweight location to calculate the gapper scale.
}
\item{For comparison, we also considered using a simple standard deviation. 
 However, the degree of background contamination was too high to give useful results in this case. As an alternative, we show results for the standard deviation estimated from \textbf{Gaussian} fits to binned data. This method has the clear disadvantages that it depends on the choice of bin (we use 200 km s$^{-1}$) and that it is not robust to the assumed underlying distribution. At the same time, it provides a visually intuitive measure of scale and an instructive comparison with the other two measures. We use an unweighted non-linear least-squares fit to the counts in bins. As with the biweight and gapper, we estimate uncertainty using bootstrap resampling.  Note that we consider a single Gaussian component rather than multiple peaks; we discuss this choice further below. To be clear, we do not assume the underlying distribution should be intrinsically Gaussian, nor do we use the results of this fit for any purpose other than as an alternative method to estimate the redshift dispersion.
}
\end{itemize}

We include a small geometrical correction factor $(\cos\alpha \cos\beta)^{-1}$ applied to the value of $cz$ for each galaxy, where $\alpha$ and $\beta$ are the latitude and longitude of the galaxy relative to the filament. This factor accounts for the fact that we cannot see the full three-dimensional velocity along our line of sight to each galaxy, if we assume a simple scenario where galaxy motions are directed toward or away from the filament. (If the mock observations included distances, so that could know the true location of each galaxy relative to the filament, we could use the more complicated correction factor described in Appendix A of \citealp{odekon2022a}.) 

Figure 2 illustrates our procedure applied to the same filament from Figure 1. The top panel shows a detailed view of the center region of the filament. 
In the top panel, gray dots indicate the values of $cz$ for individual galaxies, while blue points with error bars indicate the mean and standard deviation for overlapping segments of width 1\degree, with centers shifted by 0.4\degrees. (In our analysis we use segments shifted by 0.2\degrees; here we show half of them, so the galaxies are visible in the background.) 
The bottom three panels show the distribution of $cz$ for three sample segments: from left to right, a segment with small $\sigma$ and small $\sigma_{err}$; a segment with large $\sigma$ and small small $\sigma_{err}$; and a segment with large $\sigma$ and large $\sigma_{err}$. Large values for $\sigma_{err}$ typically result from complicated structures like infalling groups or bifurcations in the filament.

\begin{figure*}
   \includegraphics*[width=0.99\textwidth]{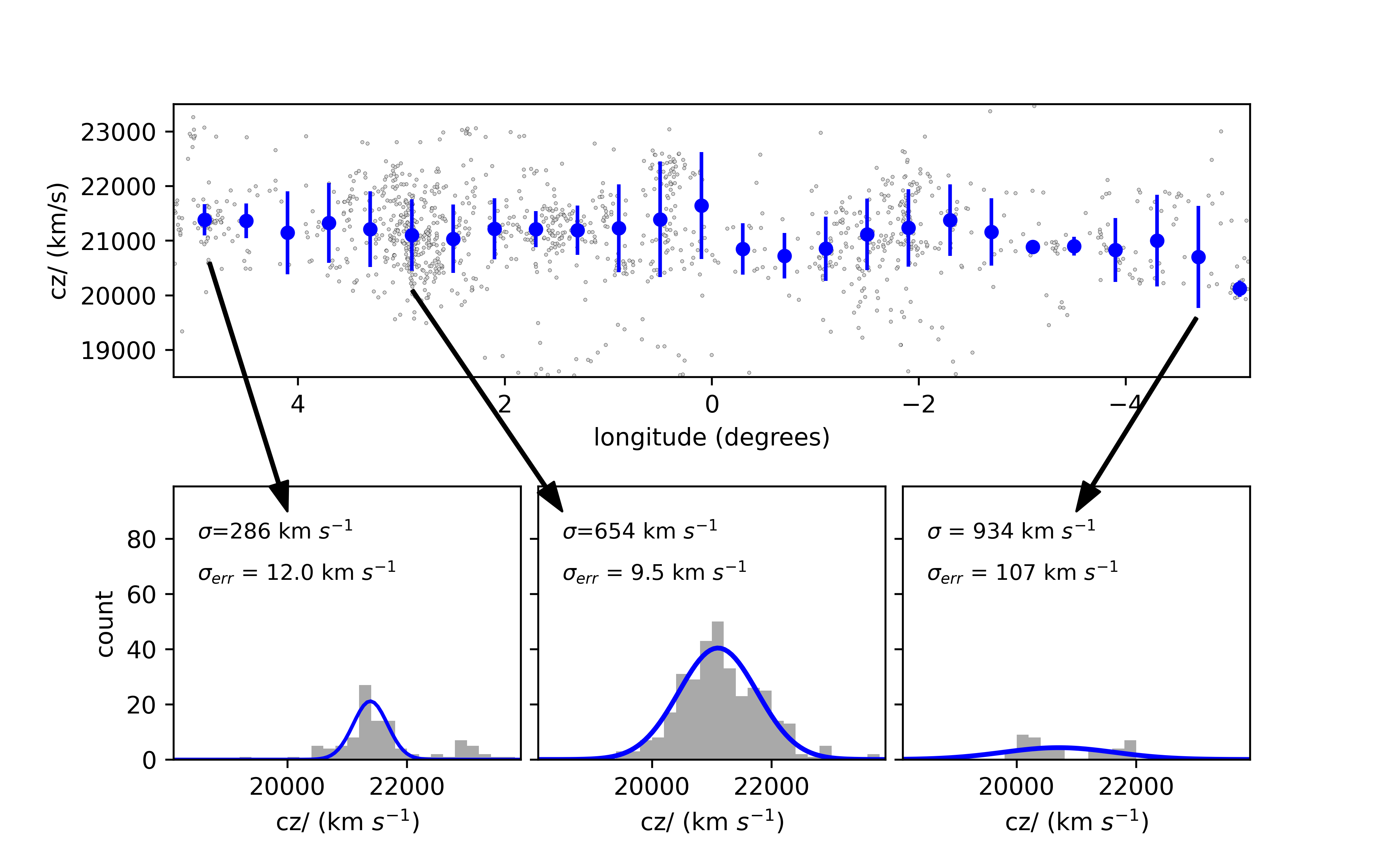}
    \caption{A detailed view of the same filament shown in Figure 1, illustrating the procedure for calculating redshift dispersions for overlapping segments along the filament. In the top panel, gray dots indicate the values of $cz$ for individual galaxies, while blue points with error bars indicate the mean $cz$ and standard deviation for overlapping segments of width 1\degree, with centers shifted by 0.4\degrees. (In our analysis we use segments shifted by 0.2\degrees; here we show half of them, so the galaxies are visible in the background.) 
    The bottom three panels show the distribution of $cz$ for three sample segments: from left to right, a segment with small $\sigma$ and small $\sigma_{err}$; a segment with large $\sigma$ and small small $\sigma_{err}$; and a segment with large $\sigma$ and large $\sigma_{err}$. Large values for $\sigma_{err}$ typically result from complicated structures like infalling groups or bifurcations in the filament.}
\end{figure*}

Before moving on to discuss how we determine the linear mass density, we pause to recognize that an option for measuring dispersion would be to use a mixture of multiple peaks. Our choice to use single-peak estimators in this context is motivated by the desire for a procedure that can be applied generally. As an example of a complication that might be dealt with on a case-by-case basis, consider a filament that bifurcates in redshift space,  creating two peaks in the redshift distribution for some particular segment. We could use a fit that takes into account both peaks, and either discard one or include them both in the analysis. But it is not straightforward to assign corresponding values for linear mass density because the dispersion tends to reflect the mass in a very large volume surrounding the axis of the filament. The two arms of a bifurcated filament may be close enough that these volumes overlap, which produces a different relationship between dispersion and mass. It is not clear from the information in redshift space how physically far apart the two arms actually are, in order to attempt to account for this. While it may be possible to find useful strategies for specific cases, our analysis here focuses on methods that can be applied more generally.

In order to find a relationship between redshift dispersion and linear mass density, we must also make a choice about how to define the linear mass density $\mu$. This would be relatively straightforward in the case of a smooth, axisymmetric mass distribution that extends out from the filament axis to a limited distance. However, supercluster-scale filaments are lumpy and embedded in a network of smaller filamentary structures. In this paper, we find the linear mass density in a cylinder of radius 10 Mpc from the axis of peak density for the segment. This volume is large enough that the results are not sensitive to the specific method for determining the peak density; for example, the results are essentially unchanged if we find the peak using the observable galaxy distribution instead of the mass distribution. 

\subsection{The Relationship Between Dispersion and Mass}
 
In order to infer the linear mass density from the dispersion, we would like to see a tight correlation between these variables. Figure 3 shows the dependence of $\log\mu$ on $\log\sigma$ obtained from the procedure described above. The top row shows the distribution for all 13200 filament segments, using each of our three measures for the dispersion, as well as the 
the analytical model of \cite{eisenstein1997a}, shown as a dotted line in all six panels in the form $\log\mu = m\log\sigma+b$, where $m=2.00$ and $b=8.67$, as determined from Equation 1.
As seen in the top row, each of the three measures of dispersion --- especially the Gaussian --- has a large amount of scatter. There are also specific concentrations of values that can be seen below the dotted line. For example, the central contours for the gapper show a concentration extending straight downward from the center, and the Gaussian has a distinct peak in concentration on the lower right.
This concentration of values, at about $\log(\sigma$/\kms)=3.0 and $\log(\mu$/\msun Mpc$^{-1}$)=13.3, indicates segments where the estimated dispersions are especially large for their masses. Inspecting the distribution of redshifts for these cases, we find that they do not have well-defined peaks in redshift; they have redshifts scattered throughout the $18,000 - 24,000$ \kms \ range we use for the dispersion estimates.  In these cases, the estimates for $\sigma$ are large because they are essentially showing the scale of the entire $18,000 - 24,000$ \kms \ range rather than the scale for a peak within that range. Indeed, if we increase the range, these estimates for $\sigma$ increase as well.

Luckily, it is possible to recognize segments for which the calculated dispersion is likely to be poor indicator of mass, using only information available in redshift surveys. For example, the bootstrap uncertainty $\sigma_{err}$ is often (although not always) quite large in these cases. We can also employ statistics that measure whether the redshift distribution is close to Gaussian --- for example, the Shapiro-Wilk statistic \citep{shapiro1965a,razali2011a}. While we do not expect the distribution to be truly Gaussian, redshift distributions for which the calculated dispersion is a poor indicator of mass are typically far from Gaussian. 

The second row of Figure 3 shows the relationship between dispersion and linear mass density when we eliminate segments using threshold values for $\sigma_{err}$ and the Shapiro-Wilk statistic \textit{W}, where \textit{W} is found using redshifts limited to within three standard deviations of the peak for each segment. Note that for the Shapiro-Wilk test, we use the test statistic \textit{W} rather than the associated p-value for the probability that the distribution is consistent with Gaussian. We are not trying to see which segments have a distribution that is statistically distinguishable from a Gaussian --- in this case, we would lose nearly all segments with a large number of galaxies, where it is statistically easy to see that the distributions are in fact not intrinsically Gaussian. Instead, we use low values of the \textit{W} statistic itself as a heuristic indicator that the redshift distribution is complicated enough that the measured dispersion is likely to be a poor indicator of mass. 
The specific choice of threshold values for $\sigma_{err}$ and \textit{W} is arbitrary in the sense that a stricter threshold would further reduce scatter, but also reduce the number of remaining segments. Our choice of $\sigma_{err}<70$ \kms and $W>0.98$ is based on balancing the goals of yielding a tight correlation between $\sigma$ and $\mu$, as well as keeping a significant fraction of segments. 

Applying the cuts for both $\sigma_{err}$ and \textit{W} simultaneously helps to restrict scatter for each of the three methods (biweight, gapper, and Gaussian) more effectively than using just $\sigma_{err}$ or \textit{W} alone. However, we note that using $\sigma_{err}$ alone is more effective in reducing scatter for the Gaussian method, and using \textit{W} alone is more effective for the biweight and gapper.

It is somewhat surprising that the analytical model of \cite{eisenstein1997a} matches the distributions in the second row of Figure 3 so well.  This model is based on the assumption of galaxy orbits that have relaxed within the large-scale gravitational potential of the filament. The complicated structure of the simulated filaments suggests that this is not the case, a topic we explore in section 4. In addition, the values for $\mu$ depend on our choice of the region within which to determine the mass --- here, a cylinder of radius 10 Mpc.  A different choice would alter the relationship between dispersion and linear mass density. We discuss this in more detail below as well.

We emphasize that recognizing which segments are ``good" is not simply a matter of selecting values with less random scatter. In particular, the selection criteria primarily eliminate segments where the estimated dispersion is systematically too high, producing specific concentrations of values that can be seen below the dotted lines in the top row of Figure 3 --- for example, the contours for the gapper extending straight downward from the center, and the concentration on the lower right for the Gaussian.

The distributions in the bottom row of Figure 3 can provide specific numbers for the inferred linear mass density for a specific measured dispersion, along with uncertainties in the inferred linear mass density. Figure 4 show these ``good" segments as light blue points. Median values of $\log\mu$ for a given $\log\sigma$ are shown as dark blue points with errors bars determined by bootstrap resampling. To facilitate estimating $\log\mu$ from $\log\sigma$, we provide a simple quadratic equation $y=a_o+a_1x+a_2x^2 $ that approximates these median values, shown in Figure 4 as solid blue lines. We first tried a simpler straight-line fit, but found they did not adequately match the median points, producing reduced $\chi^2$ values of 1.9, 3.3, and 3.4 for biweight, gapper, and Gaussian methods, respectively. The reduced $\chi^2$ values for the quadratic fit are of 1.8, 2.0, and 1.4.

Confidence intervals for the determination of $\log\mu$ for a given $\log\sigma$ are shown in Figure 4 as dark blue dashed lines.  These are based on the vertical scatter above and below the median. Specifically, they show the 16 and 84 percentiles (corresponding to 68\% of the values, or one standard deviation). 
For example, a dispersion of $\log(\sigma$/\kms) = 2.6 as measured using the biweight estimator, implies an expected linear mass density of $\log(\mu$/\msun Mpc$^{-1}) \sim$ 13.9, with an uncertainty of about $\pm 0.2$ dex. 

The first three rows of Table 1 list the parameters from the quadratic fit to the median values of $\log\mu$ for a given $\log\sigma$, along with this vertical scatter about the median, determined as the average difference between the median and the dashed lines. We caution that the values of the fits to the medians are determined only over the range shown in Figure 4; beyond this range, we do not have adequate information to assess if the quadratic function would be a good fit. Similarly, the values for the scatter are poorly determined outside the range where the dashed lines are shown in Figure 4.

A limitation to our procedure is that it cannot be applied to all segments. to some extent, this is an intrinsic limitation based on the complicated dynamics within certain regions along the filament. Nonetheless, we envision several refinements that could improve the number of segments for which we can estimate the mass. Here we explore one example: a more aggressive sigma clipping for the biweight and gapper methods, that helps especially with segments that have low masses.  A particular challenge is estimating dispersion in cases where the redshift distribution is not clearly peaked, but spread throughout the 6000 \kms window around the filament, as one might expect for a relatively low-mass section of a filament. As described above, two ways we assist the biweight and gapper estimators in finding a subtle peak at the filament are 2.5-sigma clipping and using a narrower 2000 \kms window to find the peak in the first iteration.  Nonetheless, the estimators fail for most lower-mass segments -- that is, they provide a value for the dispersion that is higher than we would expect for their mass, and we can recognize that this is likely to be true because they do not make the cut for $\sigma_{err}<70$ \kms and $W>0.98$.  We find that applying a more aggressive 2.0 sigma clipping succeeds on a significant number of those segments, providing an alternative method for cases for which the 2.5 sigma clipping fails. On the other hand, 2.0 sigma clipping produces more scatter in the relationship between $\log\sigma$ and $\log\mu$. Therefore, we recommend it for use in special cases rather than as a replacement for 2.5 sigma clipping. The last two lines of Table 1 give the values for 2.0 sigma clipping. Note the larger scatter in this case. 

Figure 5 illustrates the reconstruction of linear mass density for a simulated filament from dispersion measurements, using the information in Table 1. Each panel shows the known linear mass density (gray curve) and reconstructed linear mass density (points with error bars) for the same filament shown in Figures 1 and 2.  Dark blue points give values based on 2.5 sigma clipping, while wide light green points give values based on 2.0 sigma clipping for segments that did not qualify as ``good" with 2.5 sigma clipping.  The vertical extent of the error bars corresponds to the ``scatter'' column in Table 1, and the horizontal extent of the caps on the error bars corresponds to the physical extent of the segment. (Recall that segments have a width of 1\degree, or approximately 5 Mpc at the filament distance of 300 Mpc, and that segments are shifted by 0.2\degrees, or approximate 1 Mpc.) As illustrated in Figure 5, regions with higher density are easier to reconstruct, but some lower-density regions are also possible, especially if we include the results from 2.0 sigma clipping. 
For this particular filament, 50\% of the segments qualify as ``good"; taking into account the fact that the segments overlap, 84\% of this filament is part of at least one of these ``good" segments. More generally, for all 264 filaments together, 40\% of the segments qualify as ``good"; taking into account the overlap, 68\% of all filaments are part of at least one of these segments.

One of the segment mass estimates visible in Figure 5 is significantly off from the true value, and it is instructive to examine the reason. In the top two panels, the dark blue error bar just left of center, at about 1\degree \  (5 Mpc) is three standard deviations higher than the actual value. Examining the redshift distribution and the actual positions of the galaxies, we can see that the region between 0 and 1\degree \  includes a background group that is connected to the main filament by a smaller filamentary structure, visible in the right panel of Figure 1 just left of center at about 310-330 Mpc. This structure is falling towards the filament, and its motion shifts the values of $cz$ lower, similar to those for galaxies in the filament. However, the structure is far enough away from the filament that its mass is not included in the linear mass density shown by the gray curve.  This case offers a clear example of how the structure of filaments and their surroundings complicate the process of relating dispersion to mass.

\begin{figure*}
   \includegraphics*[width=0.99\textwidth]{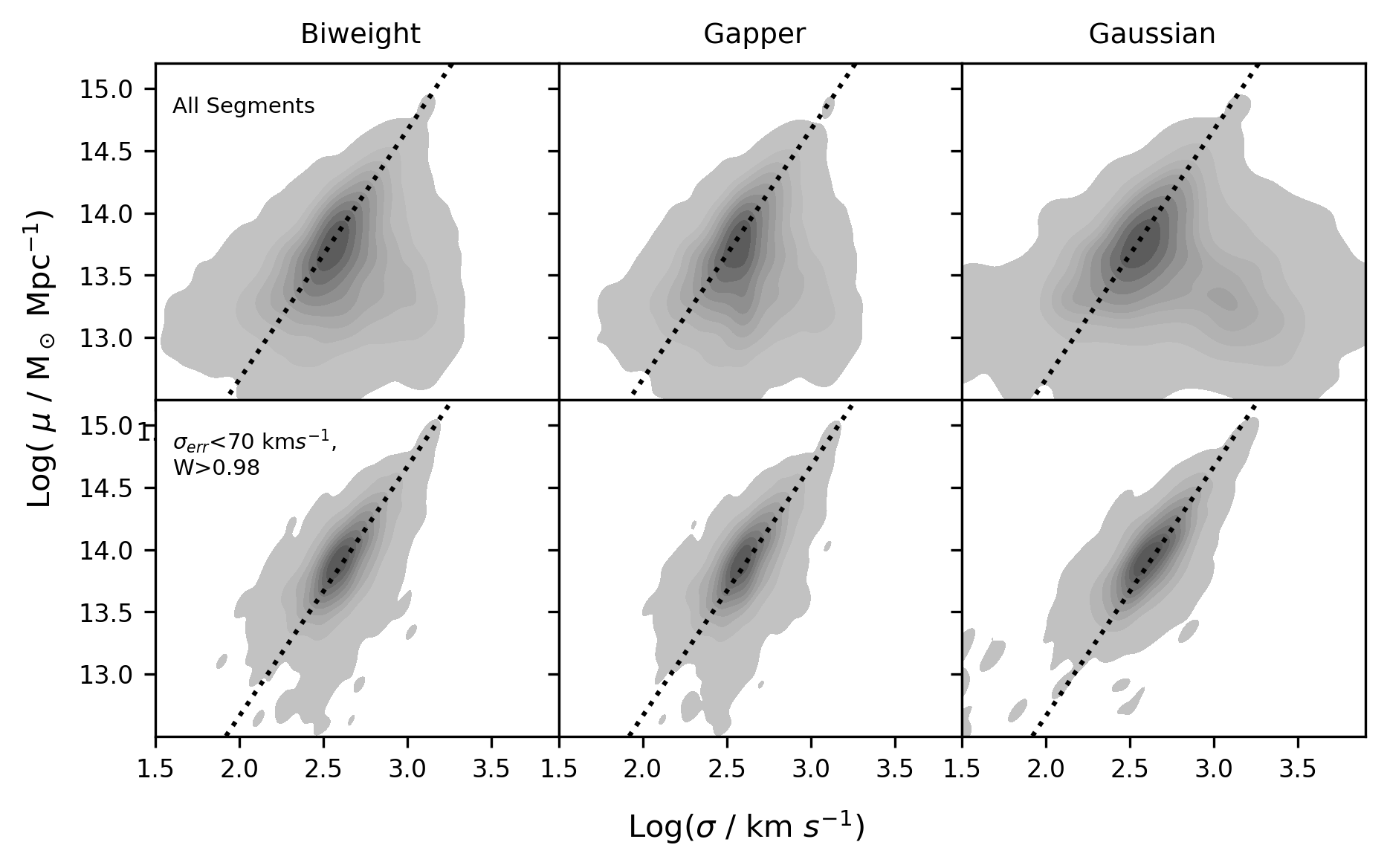}
    \caption{The relationship between redshift dispersion $\sigma$ and linear mass density $\mu$ where $\sigma$ is determined using (left to right) the biweight estimator, the gapper, or a Gaussian fit. The gray contours in the top row show the distribution for all 13200 segments, while those in the bottom row show the segments identified as ``good" based on low bootsrap uncertainty $\sigma_{err}<70$ \kms and high Shapiro-Wilk statistic $W>0.98$. From left to right, the number of segments included in the bottom row is 3624, 4245, and 2699. The dotted black line in each panel shows the result from the analytical model of \citet{eisenstein1997a}. The relationship between dispersion and linear mass density is much tighter for the ``good" segments, allowing us to use $\sigma$ as a predictor for $\mu$. It is also similar to the prediction from the analytical model.}
\end{figure*}

\begin{figure*}
   \includegraphics*[width=0.99\textwidth]{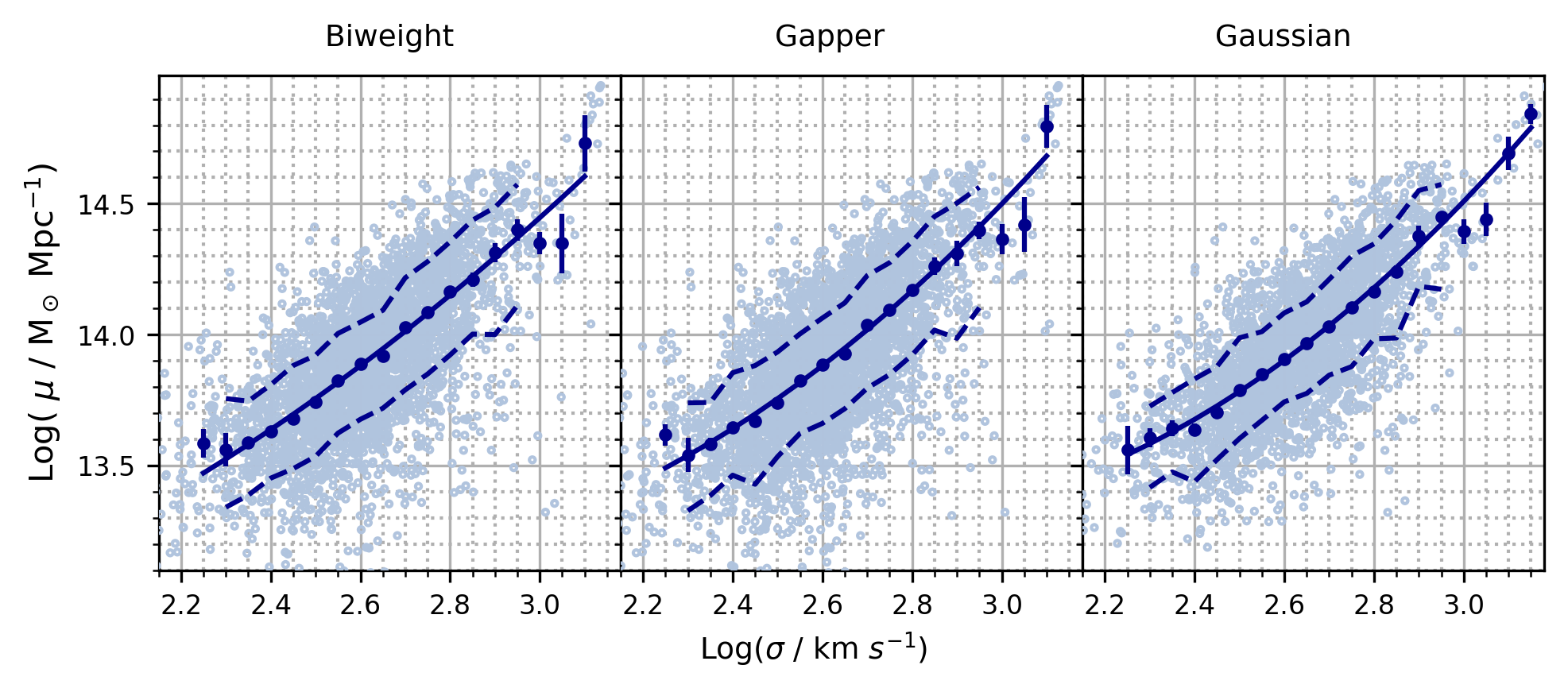}
    \caption{Median values and confidence intervals for the inferred value of the linear mass density based on the ``good" segments in Figure 3.  Dark blue points show the median values of $\log\mu$ for a given $\log\sigma$ with error bars from bootstrap resampling, and dashed lines show 16 and 84 percentiles (corresponding to 68\% of the values, or one standard deviation). Solid dark blue lines show fits to the median points using a quadratic function. Table 1 shows the parameters for these fits, as well as the scatter given by the vertical distance between the median value and the dashed lines.}
\end{figure*}

\begin{deluxetable}{lrrrrr}
\tabletypesize{\footnotesize}
\tablewidth{0.5\textwidth}
\tablenum{1}
\tablecaption{Parameters from Figure 4 \label{tab:Fits}}
\tablecolumns{7}
\tablehead{
\colhead{case} &
\colhead{a$_0$} &
\colhead{a$_1$} &
\colhead{a$_2$} &
\colhead{scatter}
}
\startdata 
Biweight     & 12.84 & -0.49 & 0.34  & 0.20 \\  
Gapper       & 14.91 & -2.11 & 0.66  & 0.21 \\
Gaussian     & 15.02 & -2.11 & 0.65  & 0.18\\  
\hline
Biweight 2$\sigma$    & 16.22 & -3.04  & 0.84  & 0.22\\  
Gapper 2$\sigma$      & 14.95 & -1.99  & 0.62  & 0.25 \\ 
\enddata
\tablecomments{Parameters are for the quadratic fits in Figure 4, where  $y=a_o+a_1x+a_2x^2 $. The scatter gives one standard deviation confidence intervals for the determination of $\log\mu$ from an observed value of $\log\sigma$. 
}
\end{deluxetable}	

\begin{figure*}
   \includegraphics*[width=0.99\textwidth]{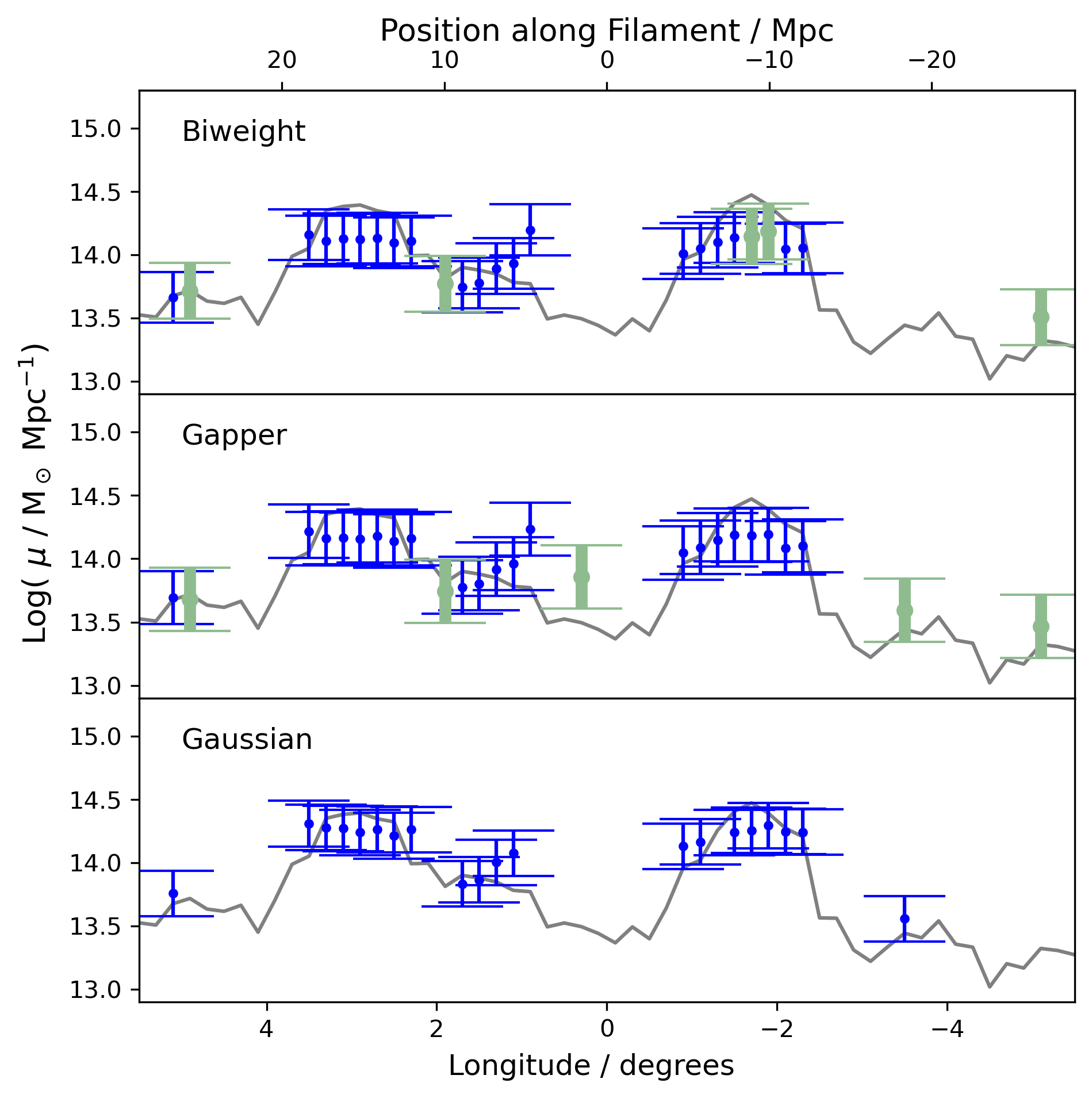}
    \caption{Reconstruction of the linear mass density for the simulated filament in Figure 1 based on the values in Table 1. Gray lines show the actual mass density, while points with error bars show the values obtained from redshift dispersions, as measured using (from top to bottom) the biweight estimator, the gapper, or Gaussian fits. The vertical extent of the error bars corresponds to the ``scatter'' column in Table 1, and the horizontal extent of the caps on the error bars corresponds to the 1$\degrees$ (5 Mpc) segment length. Dark blue points are calculated using the top three lines of Table 1, while bold light green points are calculated using the bottom two lines of Table 1 for segments that qualified as ``good" for 2.0 sigma clipping but not for 2.5 sigma clipping. The inclusion of 2.0 sigma clipping helps provide mass estimates for regions with lower densities.}
\end{figure*}

\subsection{Dependence on the Region used to Determine Mass}

It is important to recognize that the relationship between dispersion and linear mass density depends on the region within which we determine the mass. Recall that we have chosen to use the mass within a cylinder of radius 10 Mpc from the axis of peak density for the segment. To illustrate how the relationship between dispersion and mass depends on this choice, we compare the results from our cylinder with 10 Mpc radius with the results from cylinders with radii of 5 Mpc and 15 Mpc (Figure 6). In each panel, we show the results for the biweight estimator, overlaid for with the median values for the 10 Mpc case to facilitate comparison. Clearly the results depend on the region used to determine the linear mass density. While this is not surprising, since we can see that filaments have complicated structures and are not isolated, it is an important reminder that the values in Table 1 provide information about the mass as determined with a specific region.
For example, we can see that for segments with relatively low linear mass densities, changing the radius makes a large difference.  In the those cases, the volume density does not significantly decrease as the radius increases, so the linear density increases by a large fraction, and is especially sensitive to the presence of surrounding structures.   

The choice of 10 Mpc is slightly better than 5 Mpc or 15 Mpc in the sense that the $R^2$ coefficient of determination is larger. For example, the $R^2$ coefficient when we fit a quadratic to the distribution of values for all the segments yields 0.33 for 5 Mpc, 0.44 for 10 Mpc, and 0.42 for 15 Mpc. If we continue to still larger volumes, the scatter continues to increase, yielding $R^2$ coefficient of 0.40 for 20 Mpc and 0.37 for 25 Mpc.
 
Note that in our mock observations, the galaxies used to determine $\sigma$ are selected by redshift rather than distance. One might wonder whether the galaxies used to determine $\sigma$ are themselves primarily within 10 Mpc. We address this question in Section 4 below.

\begin{figure*}
   \includegraphics*[width=0.99\textwidth]{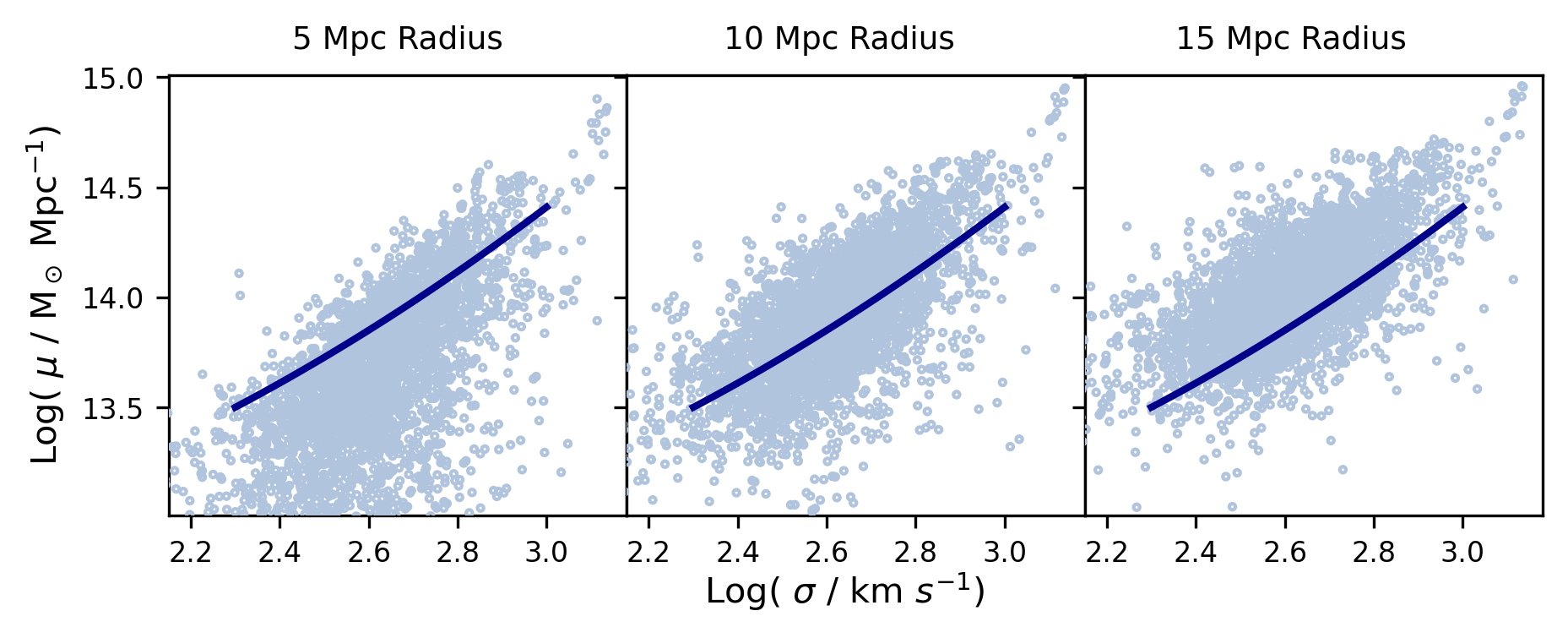}
    \caption{The relationship between $\log\sigma$ and $\log\mu$  when $\mu$ is found using cylinders with radii of 5 Mpc, 10 Mpc, and 15 Mpc, and $\sigma$ is found using the biweight estimator. To facilitate comparison, the fit to the median values for the 10 Mpc case is overlaid in each panel. Because filaments are not isolated and do not have well-defined boundaries, there is not an absolute relationship between $\log\sigma$ and $\log\mu$; it depends on the region used to define $\mu$.} 
\end{figure*}
\
\subsection{Dependence on Distance to the Filament}

In this paper, we adopt 300 Mpc as our fiducial filament distance, and $r<19.5$ as our magnitude limit. In order to be useful for actual filaments, however, our method -- and ideally, the specific values in Table 1 -- should be robust to a range of filament distances. In particular, we wish to be sure that differences in viewing angle and the shape of the observing volume for different filament distances do not change our results.

For example, in our fiducial case we select a $10\degree \times 2\degree$ wedge, corresponding to approximately 53 Mpc $\times$ 10 Mpc at the distance of the filament.  To view approximately the same physical volume from a distance of 75 Mpc -- similar to that for the well-studied Pisces-Perseus Supercluster (PPS) filament \citep[e.g.,][]{haynes1986a,odonoghue2019a} -- we would select a much larger $40\degree \times 8\degree$ wedge. This implies a much greater range in viewing angle relative to the filament, affecting the components of velocity and position that we observe. (Alternatively, we could view a smaller physical volume surrounding the filament, but selecting galaxies in a smaller physical volume might also affect our results, and would limit us to smaller range along the filament.)
 
In addition, the volume of the wedge is larger on the far side of the filament than on the near side, which will tend to skew the distribution of galaxy distances to larger values. If we consider a simple case where redshift is proportional to distance (that is, where there are no peculiar motions), then the distribution of redshift will be also skewed to larger values, potentially affecting our estimate of both location and scale.  In our case, we expect peculiar motions of galaxies near a filament to be strongly affected by infall, so that peculiar motions of the more distant galaxies produce a lower redshift, while those of the closer galaxies produce a higher redshift, skewing the redshift distribution toward lower values. These considerations matter when considering different filament distances because the relative volume in the foreground and background will be especially different for nearby filaments, for which the opening angle of the wedge is large.

Figure 6 shows the relationship between dispersion and mass for four filament distances when we change the opening angles to subtend the same physical size at the distance of the filament.  In addition to examining cases $\pm 150$ Mpc from our fiducial distance, we examine the even closer distance of 75 Mpc in order to include a case similar to the Pisces-Perseus Supercluster (PPS) filament. For each distance, we use segments with the same physical length of 5 Mpc and same 
\textit{absolute} limiting magnitude of about -17.9  as in the fiducial case.  Note that this makes it easier to perform this analysis on closer filaments, in the sense that we do not need to reach an apparent magnitude as faint as $r=19.5$.  For example, if the PPS filament at about 75 Mpc were in a part of the sky covered by the of the Sloan Digital Sky Survey spectroscopic survey, the magnitude limit of r=16.5 would be well within the SDSS main galaxy sample limit of $r \sim 17.77$ \citep{stoughton2002a}.

If instead of using a fixed absolute magnitude, we maintain the same apparent magnitude limit of $r<19.5$ for closer filaments, the higher number of galaxies included tends to produce slightly higher values for $\sigma$, especially for segments with relatively small masses. It is not surprising that this might be the case, since fainter galaxies are more evenly distributed and can fill in the spaces between peaks in redshift. Similarly, if we keep the same apparent segment length of 1\degree on the sky, instead of the same physics segment length of 5 Mpc, there is a slight shift in the relationship between dispersion and linear mass density. We do not find this surprising, because the dispersion and the criteria to select ``good" segments are based on redshifts in regions that are dynamically different - for example, regions that combine segments that pass the criteria for ``good" for 5 Mpc segments may be combined with regions that did not pass the criteria for 5 Mpc segments. 

\begin{figure*}
   \includegraphics*[width=0.99\textwidth]{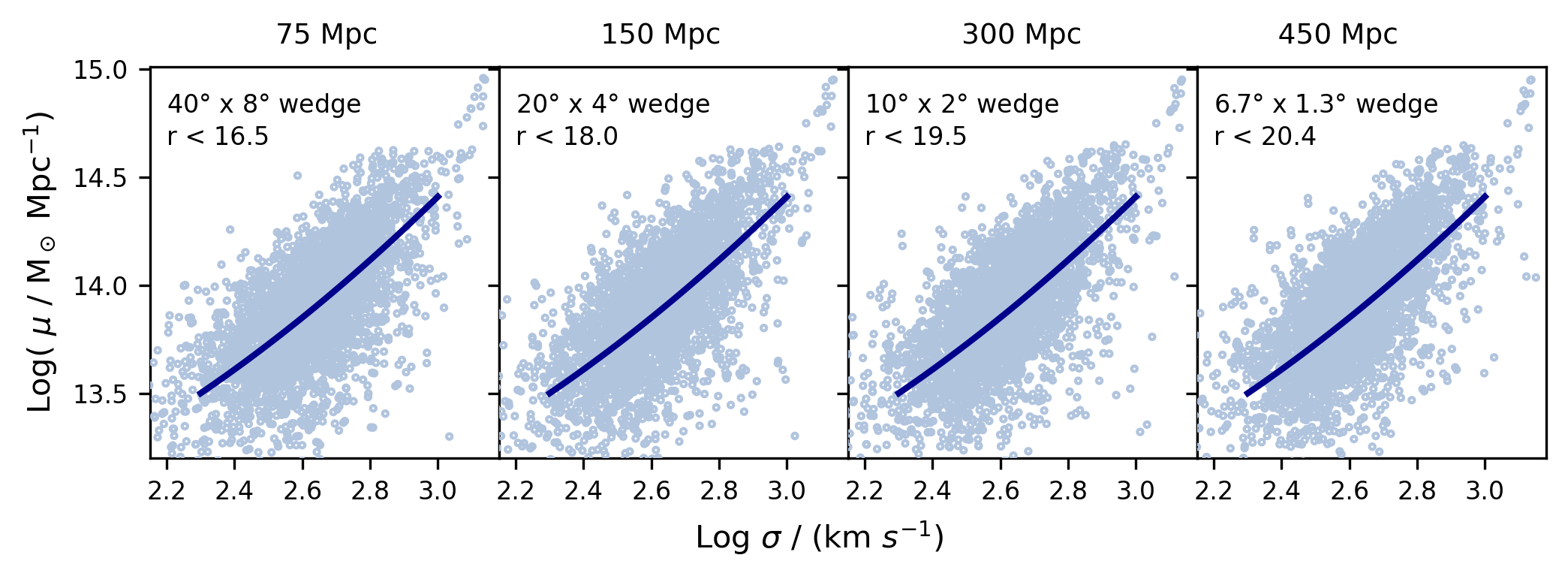}
    \caption{
    The relationship between $\log\sigma$ and $\log\mu$ when the filaments are placed at distances ranging from 75 Mpc to 450 Mpc, and $\sigma$ is found using the biweight estimator. To facilitate comparison, the fit to the median values for the 300 Mpc case is overlaid in each panel. The wedge shape and apparent magnitude limit are given in the upper left of each panel. The relationship found for the 300 Mpc case holds across this wide range of distances if the absolute magnitude cutoff and the physical segment size are held constant, despite the range of wedge shapes.}
\end{figure*}


\section{The Dynamical State of Galaxies That Appear in Filaments}

As shown in the bottom row in Figure 3, the results for mock observations of our ``good" segments are similar to the expectations from the analytical model of \cite{eisenstein1997a} that assumes dynamical equilibrium. At the same time, it is reasonable \textit{not} to expect the results of simulations to match a simple model of isolated filaments in dynamical equilibrium, given their complicated structure. And as illustrated in Figure 6, filaments are not isolated and do not have well-defined boundaries, so there are multiple ways to define the corresponding mass; changing the region used to find the mass changes the relationship between redshift dispersion and linear mass density.

In this section, we examine the possibility that the filaments are in dynamical equilibrium. In analogy with a simple model for clusters of galaxies, one could picture a roughly cylindrical virialized region, where galaxies are, on average, no longer falling toward the filament, but have orbits that reflect the large-scale gravitational potential of the filament. From the information in a redshift survey alone, it would not be clear whether this is the case. In particular, there are two well-known effects from redshift-space distortions that are relevant to supercluster-scale filaments \citep{kaiser1987a}. One is the elongation of galaxy redshifts in groups and clusters caused by orbits within the gravitational potential.  In the context of supercluster-scale filaments, groups and clusters are substructures, where internal orbits do not reflect the gravity of filament as whole.  The second type of redshift-space distortion is the tendency for infall toward large-scale structures like filaments and walls to make them look thinner in redshift space. For a filament oriented across the line of sight, the infall of background galaxies gives them a negative peculiar velocity that counteracts the larger expansion redshift caused by its larger distance.  Similarly, the infall of foreground galaxies imparts a positive peculiar velocity that counteracts the smaller expansion redshift caused by its smaller distance.  This creates the triple-valued redshift that is also well-known in the context of infall to clusters: at the average redshift of the cluster or filament, there can also be background infall and foreground infall.

Figures 8 and 9 illustrate the dynamics of galaxies within the mock observation for the same filament shown in Figure 1, split into galaxies in halos with mass $< 10^{13}M_\odot$ (``isolated galaxies") and those in halos with mass $>10^{13}M_\odot$  (``cluster galaxies"). Figure 8 shows the results for the ``isolated" galaxies. 
The left panel shows the redshift distribution for all galaxies in the mock observations (gray) and for those in the limited range 18000-24000 \kms (blue). The center panel shows the line-of-sight peculiar velocities for the galaxies in that limited range, which would appear to be ``in the filament" based on redshift. 
In the panel on the right, peculiar velocities are plotted as a function of their position along the filament, with color indicating the distance to each galaxy. Unlike the distribution of redshifts, the distribution of peculiar velocities is bimodal. Galaxies with negative peculiar velocities tend to be at larger distances (colored light yellow), while galaxies with positive peculiar velocities tend to be at smaller distances (colored dark purple).  In other words, most isolated galaxies identified as ``in the filament" based on their redshift are actual falling toward the filament. Figure 9 shows the same process applied to galaxies in halos with mass $>10^{13}M_\odot$. This case shows the affect of orbits within substructure; in the panel on the right, individual clusters show as vertical ``fingers" of galaxies at approximate the same longitude and approximately the same distance (same color). Of all the galaxies in the mock observation of the filament and in 6000 \kms window around the filament, 56\% are in halos with mass $>10^{13}M_\odot$, while the remaining 44\% are in the ``isolated" category where motions are dominated by infall. 

When we examine the velocities and distances of galaxies (not just redshifts), we see that the relationship between redshift dispersion and mass --- despite its similarity to the results from an analytical model that assumes equilibrium within the potential of the filament as whole -- is strongly affected by both infall and orbits within substructures.

A traditional way to see the effect of infall is through the relationship between peculiar velocity and distance. This is shown in Figure 10 for the same sample filament. Points show galaxies from the mock observations over the distances 244-345 Mpc. Infall is evident as positive peculiar velocities on the near side of the filament and the negative peculiar velocities on the far side. Our $cz$ selection range 18000-24000 \kms\  is indicated by slanted dashed lines, and galaxies in that range are color-coded by $cz$, as shown in in the inset histogram. We can see that most of the infall region is included in the selected range of $cz$.

Vertical lines on Figure 10 indicate distances within 10 Mpc of the filament, approximately the region used to determine $\mu$.  If our analysis were based on the full information available in the simulations, as opposed to mock redshift surveys, it might be a natural choice to use the same volume for galaxy selection (to determine $\sigma$) and for the determination of $\mu$. Using our selection criteria, we find that for the filament in Figure 10, 70\% of the galaxies used to determine $\sigma$ based on their redshift are within 10 Mpc of the filament distance. (For all the filaments in total, 60\% percent of the galaxies selected by redshift are within 10 Mpc of the distance of the filament.)  We can also see that while infall is visible throughout the distance range shown in Figure 10, the peak of infall --- that is, the highest deviation of average galaxy velocity for a fixed distance --- appears to be within 10 Mpc. Indeed, \cite{odekon2022a} found that the location of maximum infall toward supercluster-scale filaments is typically between 5-10 Mpc from the filament axis, suggesting that 10 Mpc might be an appropriate choice for both galaxy selection and mass selection in cases where we use the full information available in the simulations.

Another result from \cite{odekon2022a} that is relevant here is the difficulty of cleanly separating out galaxies that are falling into the cluster from galaxies that are ``in" the cluster. For example, if we define the filament spine as the principle axis of its large groups, as done here and in \cite{odekon2022a}, we do not find a well-defined region of space that is clearly virialized in the sense of having an average peculiar velocity of zero. While the peculiar velocity is zero at the filament axis itself, there is evidence of infall immediately out it.  For this reason, we do not attempt to quantify the fraction of infall galaxies in the context of these mock observations, but instead emphasize that it is clearly an important aspect of galaxy dynamics in the region of the filament.

\begin{figure*}
   \includegraphics*[width=0.99\textwidth]{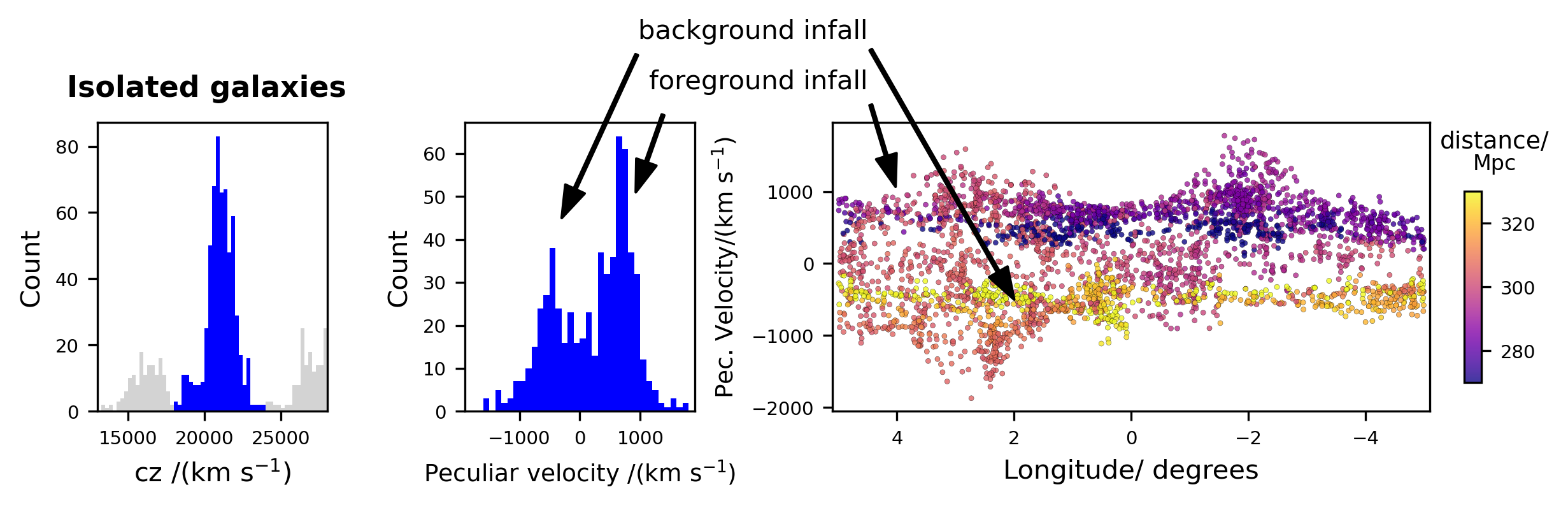}
    \caption{Illustration of the motions of galaxies in halos with $M<10^{13}M_{\odot}$, for the same filament shown in Figure 1. The left panel shows the distribution of cz for all galaxies in the mock observations (gray) and for those in the limited range 18000-24000 \kms (blue). The center panel shows the line-of-sight peculiar velocities for the galaxies in this limited range of cz, which would appear to be ``in the filament" based on redshift. The right panel shows the peculiar velocities for all galaxies as a function of their position along the filament, with distance indicated by color. The bimodal pattern of positive peculiar velocities for foreground galaxies (dark purple) and negative peculiar velocities for background galaxies (light yellow) shows that the galaxies which appear to be ``in the filament" based on their redshifts are primarily falling in from either side.}
\end{figure*}

\begin{figure*}
   \includegraphics*[width=0.99\textwidth]{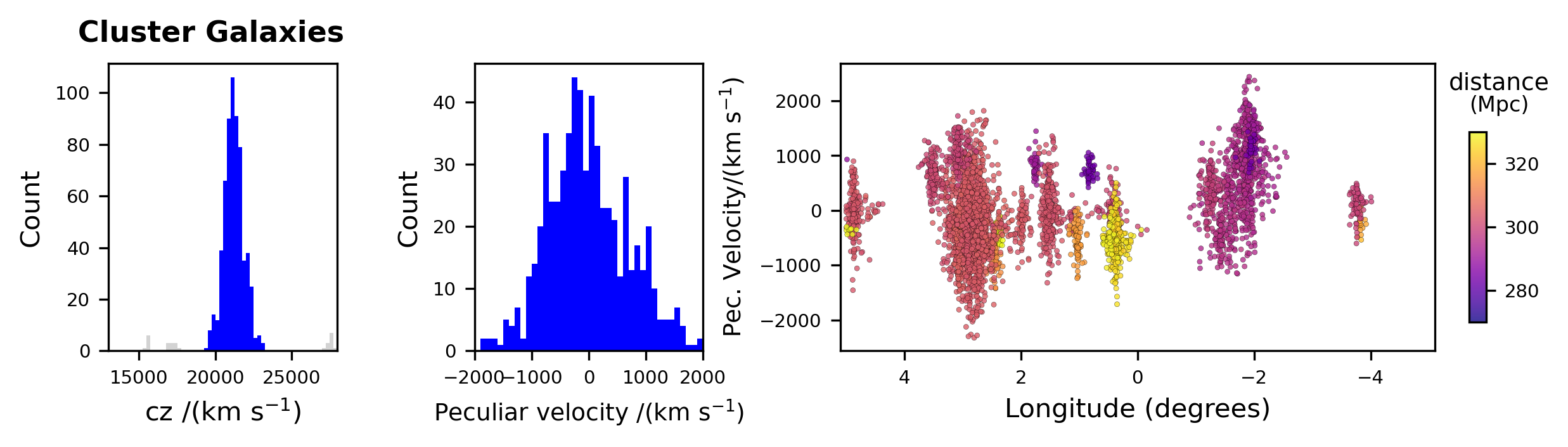}
    \caption{llustration of the motions of galaxies in halos with $M>10^{13}M_{\odot}$, in the same format as Figure 8. In the right panel, each cluster appears as a vertical band of galaxies at approximately the same longitude and approximately the same distance (color). This case highlights the presence of substructure, for which motions are driven by the orbits within individual groups and clusters.}
\end{figure*}

\begin{figure*}
   \includegraphics*[width=0.99\textwidth]{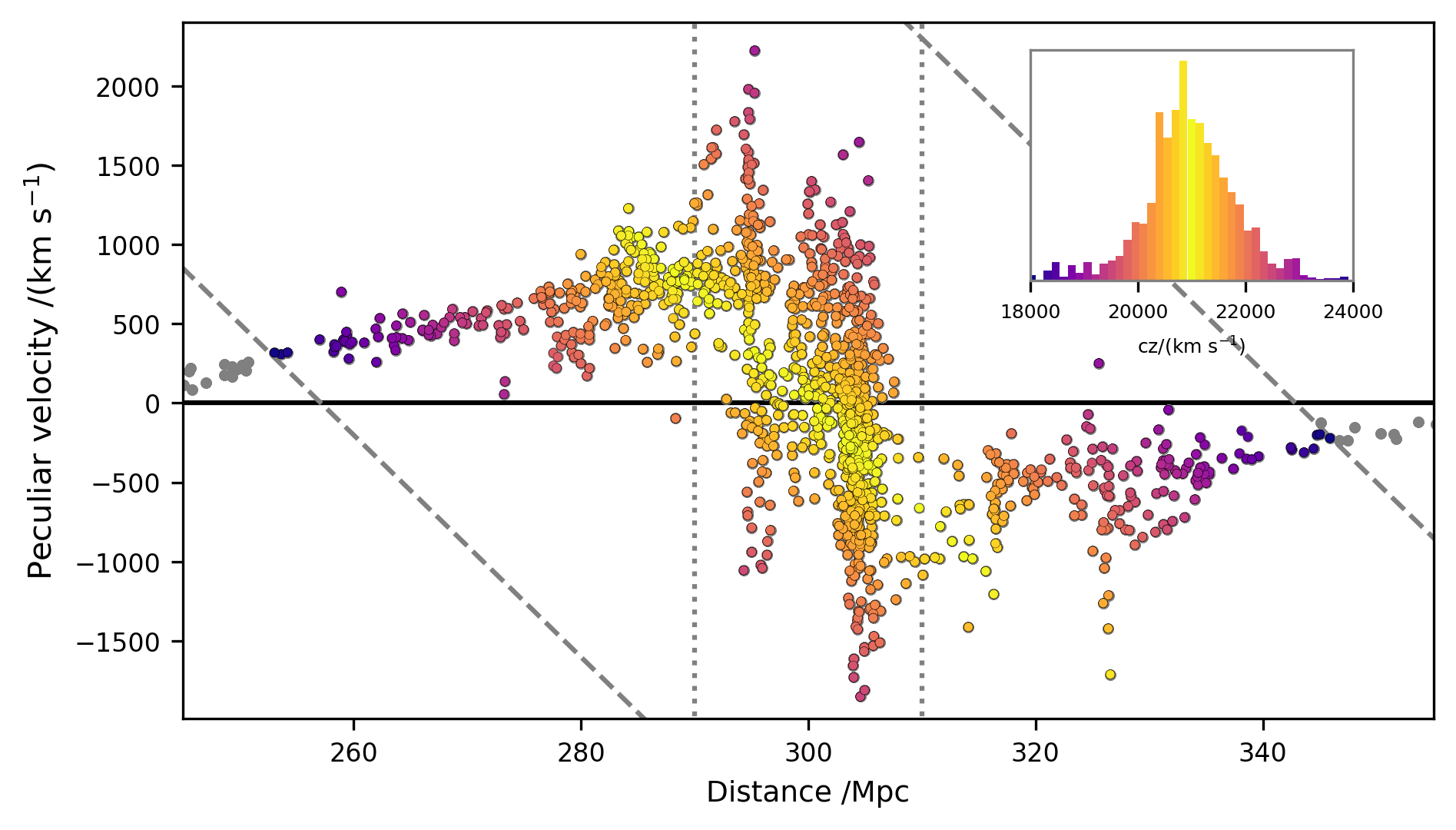}
    \caption{Infall toward same sample filament as the previous figure, as shown through the relationship between peculiar velocity and distance. Points show galaxies from the mock observations over the distances 244-345 Mpc. Infall is evident as positive peculiar velocities on the near side of the filament and the negative peculiar velocities on the far side. Our $cz$ selection range 18000-24000 \kms\  is indicated by slanted dashed lines, and galaxies in that range are color-coded by $cz$, as shown in in the inset histogram. We can see that most of the infall region is included in the selected range of $cz$.Vertical lines indicate distances within 10 Mpc of the filament distance, approximately the region used to determine $\mu$; this region includes 70\% of the galaxies in the redshift selection range.}
\end{figure*}

\section{Summary and Discussion}

We use mock observations of 264 supercluster-scale filaments to develop a strategy for estimating linear mass density based on their width in redshift space. As with dynamical mass estimates on other scales, this provides a mechanism for measuring the dark matter distribution -- or alternatively, a test for our theories of gravitation and cosmology. Our primary results can be summarized as follows:

\begin{itemize}
    \item {A challenge to finding a simple relationship between redshift dispersion $\sigma$ and linear mass density $\mu$ is the complicated structure of supercluster-scale filaments - including embedded groups and clusters, infall in the foreground and background, and bifurcations. Our approach is to consider separate, overlapping 5 Mpc segments along the filaments. It is possible to select the segments that are likely to provide a velocity dispersion that relates to their mass, using criteria for a low bootstrap uncertainty in dispersion $\sigma_{err}$ and a redshift distribution that similar to a Gaussian. Specifically, we use the selection criteria $\sigma_{err}<70$ \kms and $W>0.98$ where the Shapiro-Wilk statistic W is found within 3$\sigma$ of the peak redshift. The percentage of filament regions that fit these criteria for at least one overlapping segment is about 70\%. }

    \item {We provide equations that approximate the median value of $\log\mu$ for a measured value of $\log\sigma$ for the selected ``good" segments, along with the uncertainty in that expected value caused by the intrinsic scatter in filaments. The parameters for these equation are given in Table 1.}
    
    \item {The relationship between $\log\mu$ and $\log\sigma$ is similar to that from the analytic equilibrium model of \cite{eisenstein1997a} when we define the mass with a radius of 10 Mpc.  However, the simulated filaments are not in dynamical equilibrium; they are strongly affected by substructure and infall.}
\end{itemize}

For clarity, we note that the values in Table 1 are based on inferring the mass density based on the dispersion. This way of quantifying the relationship between mass and dispersion is different from finding the expected value of \textit{dispersion} if \textit{mass} were the known quantity, as one might do when analyzing simulations in a context other than mock observations. If the relationship between $\log\sigma$ and $\log\mu$ were simply a straight line with no scatter, these two approaches would give mathematically equivalent results in the sense that one could, for example, start with the equation $y=mx+b$ and solve for x to find the inverse relationship $x(y)$. This is not the case here, where there is a range of values for $\log\mu$ for any fixed value of $\log\sigma$, so it matters which is the dependent variable.

Our approach is to find a simple, general strategy that can be applied systematically. We envision several possible refinement that could help in providing mass estimates for segments where this method fails. We explore one such refinement here, using a more aggressive two-sigma clipping that helps provide estimates for segments with low masses. Similarly, one might use a different strategy for regions that are bifurcated in redshift space. In this case, the problem is not simply fitting to a multi-modal redshift distribution; a more fundamental issue is how to define the corresponding mass. One approach could be to determine the mass in a narrower cylinder and relate that mass to the dispersion from a multi-modal analysis of the redshift distribution.

The dynamical mass estimates in this paper apply to a scale intermediate between individual clusters of galaxies and the larger-scale filament infall patterns discussed in \cite{odekon2022a}. In terms of finding effective ways of measuring redshift dispersion and relating them to mass, our approach is very similar to mass estimates for individual clusters. On the other hand, supercluster-scale filaments are more complicated than individual clusters. While it is recognized that many clusters include substructure that tends to inflate dynamical mass estimates \citep[e.g.][]{old2018a,damsted2023a}, in the case of filaments it is less accurate to picture the system as a relaxed system contaminated by substructure and infall. It is more accurate to picture the system as primarily \textit{composed} of substructure and infall.

The approach of using large-scale galaxy infall patterns in \citep{odekon2022a}, differs not only in scale but also in that it requires redshift-independent distances to map out the peculiar velocity field surrounding each filament. Because redshift-independent distance are likely to have large uncertainties, that strategy involves averaging over many galaxies to find an average infall profile for the filament as whole.

In this paper we consider filaments on a specific scale and orientation: supercluster-scale filaments that are oriented across the sky. One might ask if these filaments are too large and/or too carefully oriented to be represented in the nearby universe. When we apply the same filament-finding algorithm used to identify simulated filaments to actual observations, the results are encouraging. For example, if we apply the filament finder to the 2MASS groups catalog of \cite{tully2015a}, as described in \cite{odekon2022a} in the context of the Pisces-Perseus supercluster filament, we find 18 supercluster-scale filaments in the nearby volume extending out to about 150 Mpc. Of these 18 filaments, five are oriented approximately across the sky in the sense that the clusters used to identify them are all within 300 \kms of the filament average, corresponding to about $\pm$ 4.3 Mpc if the redshift is attributed to distance.  This is much smaller than the typical filament length of about 50 Mpc, and it is comparable to the variation in cluster distances about the principal axis.  

Overall, our results present an encouraging prospect for dynamical mass estimates for supercluster-scale filaments with current and upcoming redshift surveys. We find that it is feasible to use this method out to at least 450 Mpc, if we have galaxy redshifts to a limiting magnitude of 20.4. In the fiducial case, filaments are placed at 300 Mpc and we use a limiting magnitude of 19.5, similar to that for the Galaxy and Mass Assembly survey \citep{driver2022a}, and the Dark Energy Spectroscopic Instrument Bright Galaxy Survey \citep{hahn2023a}.

\section{Acknowledgements}

We are grateful for the contributions of all members of the Undergraduate ALFALFA Team.  This work has been supported by NSF grants AST-1637339 and AST-1637271.  

\bibliography{mybib}

\end{document}